\documentstyle[11pt,newpasp,twoside,epsf]{article} 
\def\edcomment#1{\iffalse\marginpar{\raggedright\sl#1\/}\else\relax\fi} 
\reversemarginpar 

\begin{document} 
\title{Multi-wavelength observations of stellar populations in 
Galactic globular clusters}

\author{F. R. Ferraro} 
\affil{INAF--Osservatorio Astronomico di Bologna, via Ranzani 1, 40127 Bologna, 
Italy}

\begin{abstract}  I report on some recent results 
 in the framework of a complex project 
   aimed   to  characterize
  the photometric properties of 
  stellar populations in Galactic Globular Clusters.
\end{abstract}

\section{Introduction}
 
Galactic globular  clusters (GGCs) are extremely important astrophysical objects
since   {\it (i)} they are prime laboratories for testing stellar evolution;
{\it (ii)} they are ``fossils'' from the epoch of galaxy formation, and thus
important cosmological tools; {\it (iii)} they serve as test particles for
studying the dynamics of the Galaxy.
A few years ago our group started a long-term project devoted to study the
global stellar population in a sample of "proto-type" 
GGCs following  a multi-wavelength
approach: IR and optical observations to study cool giants
and UV observations to study blue hot sequences (Horizontal Branch (HB),
 Blue Stragglers Stars (BSS), etc).  In this paper I report a short 
summary of the most recent
results.

\section{Calibrating the
photometric properties of the Red Giant Branch in the IR}

The advantage of observing  cool giants
 in the near IR is well known since many
years. 
The contrast between the red giants and the unresolved background
population in the IR bands is greater than in any optical region, so
they can be observed with the highest S/N ratio also in the innermost
region of the cluster. Moreover, when combined with optical observations,
IR magnitudes provide useful observables such as   the
V--K color,  an excellent indicator of the stellar effective
temperature (T$_{e}$), and allows a direct comparison with theoretical model
predictions.
Since the peeonering work by Frogel and collaborators 
(Frogel, Cohen \& Persson 1983)
in the early 80's, many groups have performed
systematic IR
observations in (mainly) heavily-obscured GGCs
(see Frogel et al. 1995; 
Kuchinski \& Frogel, 1995; Minniti et al. 1995; Davidge
2000;
Ortolani et al. 2001). 
 
\begin{figure}
\plotfiddle{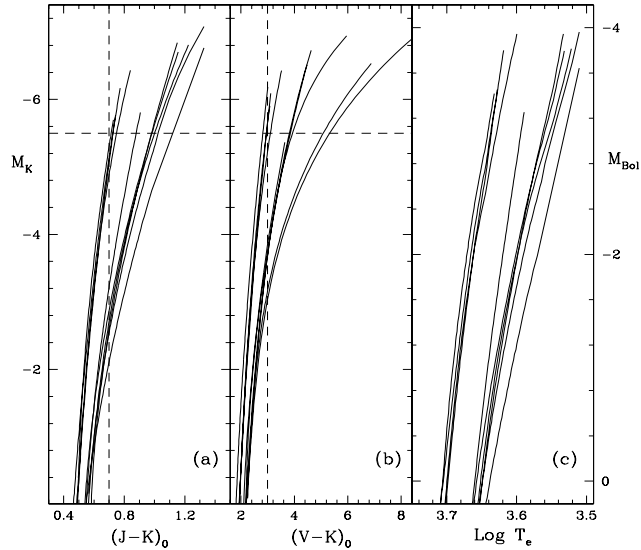}{7cm}{0}{46}{43}{-140}{-85}
\caption{RGB fiducial ridge lines for the 10 GGCs in the F00 sample in the
M$_K$,(J--K)$_0$,   
M$_K$,(V--K)$_0$
 and ($M_{Bol}, Log(T_{e})$)
  planes, ({\it panel (a)}, {\it  (b)} and {\it  (c)}), respectively.
The dashed lines indicate the magnitude levels at which some of the
parameters defined in  F00 are measured.}
\end{figure}

 In Ferraro et al. (2000, hereafter F00)   
 a new set of high quality near-IR Color Magnitude Diagrams (CMDs) 
 was presented  
  for a sample of 10 GGCs, spanning a wide range in metallicity.
We used this homogeneous data--base   to
define a variety of 
observables   allowing the  complete  characterization 
of the photometric properties of the Red
Giant Branch (RGB), namely:
{\it (a)} the location of the RGB in the CMD both in (J--K)$_0$ and 
(V--K)$_0$ colors at different absolute K magnitudes 
(--3, --4, --5, --5.5) and in temperature;
{\it (b)} its overall morphology and slope;
{\it (c)} the luminosity of the Bump and of the Tip.
All these quantities have been measured with a homogeneous procedures
applied to each individual CMD by adopting the distance moduli scale  
defined in Ferraro et al. (1999a, hereafter F99).
The mean ridge lines for the selected clusters,
 in various planes are shown in Figure 1.

A set of relations linking the photometric parameters to the cluster global 
metallicity ($[M/H]$, see F99) has been obtained in F00. Such a 
set of equations
can be very useful  to derive photometric estimates of  the metallicity
distributions 
in complex, {\it i.e.} chemically inhomogeneous, stellar populations 
 as those observed in nearby dwarf galaxies.
Indeed, one of the most puzzling (and nearest) 
examples of complex stellar population
is just in the   Halo of the Galaxy: the cluster $\omega$ Centauri.

$\omega$ Centauri is  
  the most massive globular cluster of the Milky Way ($3\times
10^6~M_{\odot}$), and 
it is the only known galactic globular which shows clear
and undisputed variations in the heavy elements content of its
giants. Recent wide field photometric surveys
(Lee et al. 1999, Pancino et al. 2000)
 have shown the existence  
of a previously unknown anomalous RGB (RGB-a).
 We have recently obtained extensive  J,K observations  
 in a wide region ($13' \times 13'$) around the cluster center.
Figure 2 {\it (left panel)} shows the CMD obtained  combining 
IR observations with the optical catalog by Pancino et al (2000).
 The mean ridge line of the dominant (metal poor) population and 
 of the anomalous RGB-a (metal rich) population are overplotted
 to the CMD, and  compared (in Figure 2-{\it left panel})  with
  the  mean ridge lines of three  reference clusters 
  (from F00).  
 The  shape of the RGB-a and its position in the
 CMD   indicate
 a   metallicity of $[M/H]\sim-0.5$,
 which is fully consistent with the most 
  recent direct spectroscopic determination  
 (Pancino et al 2002).

\begin{figure}
\vspace{-0.in}
\plottwo{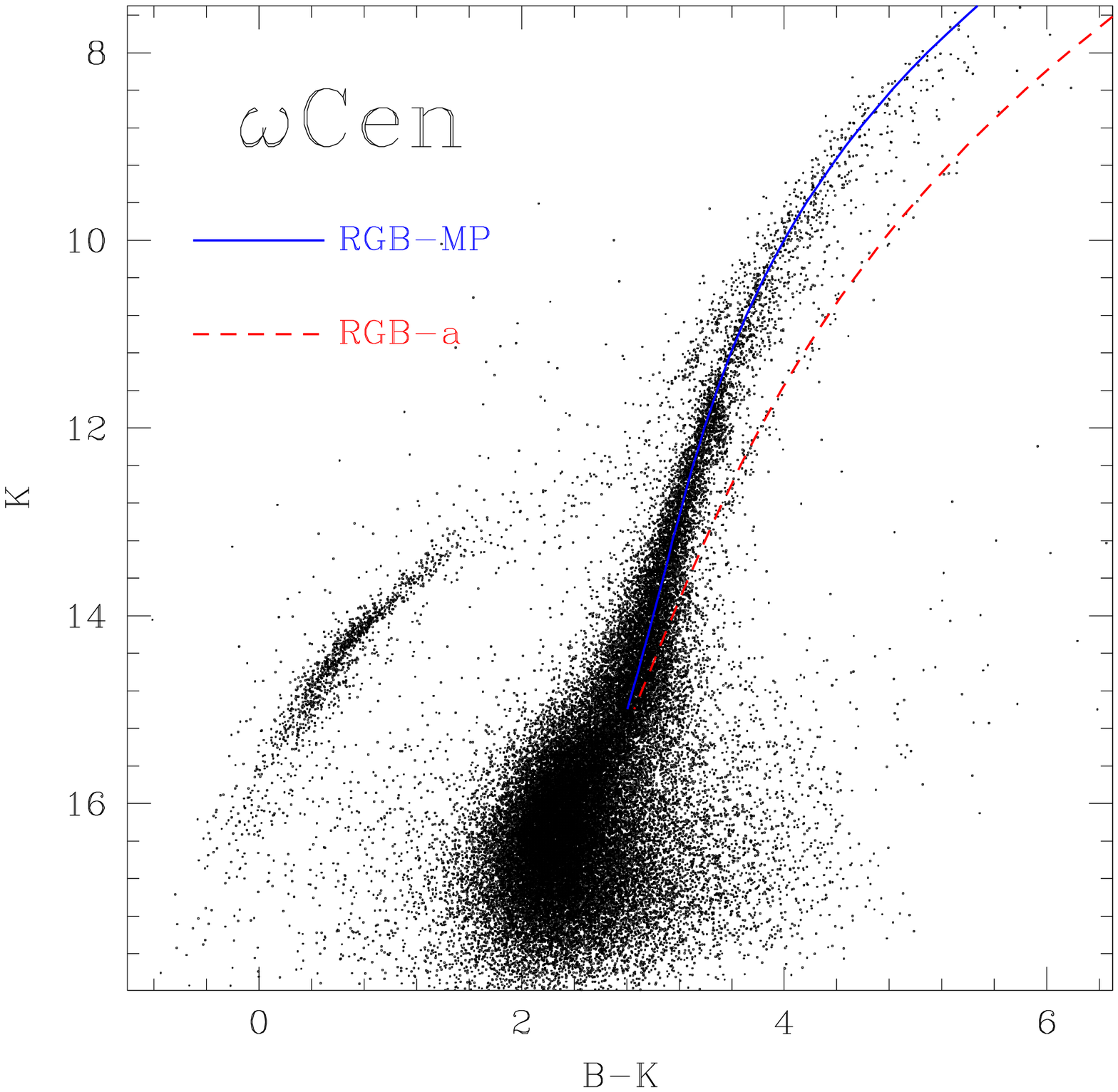}{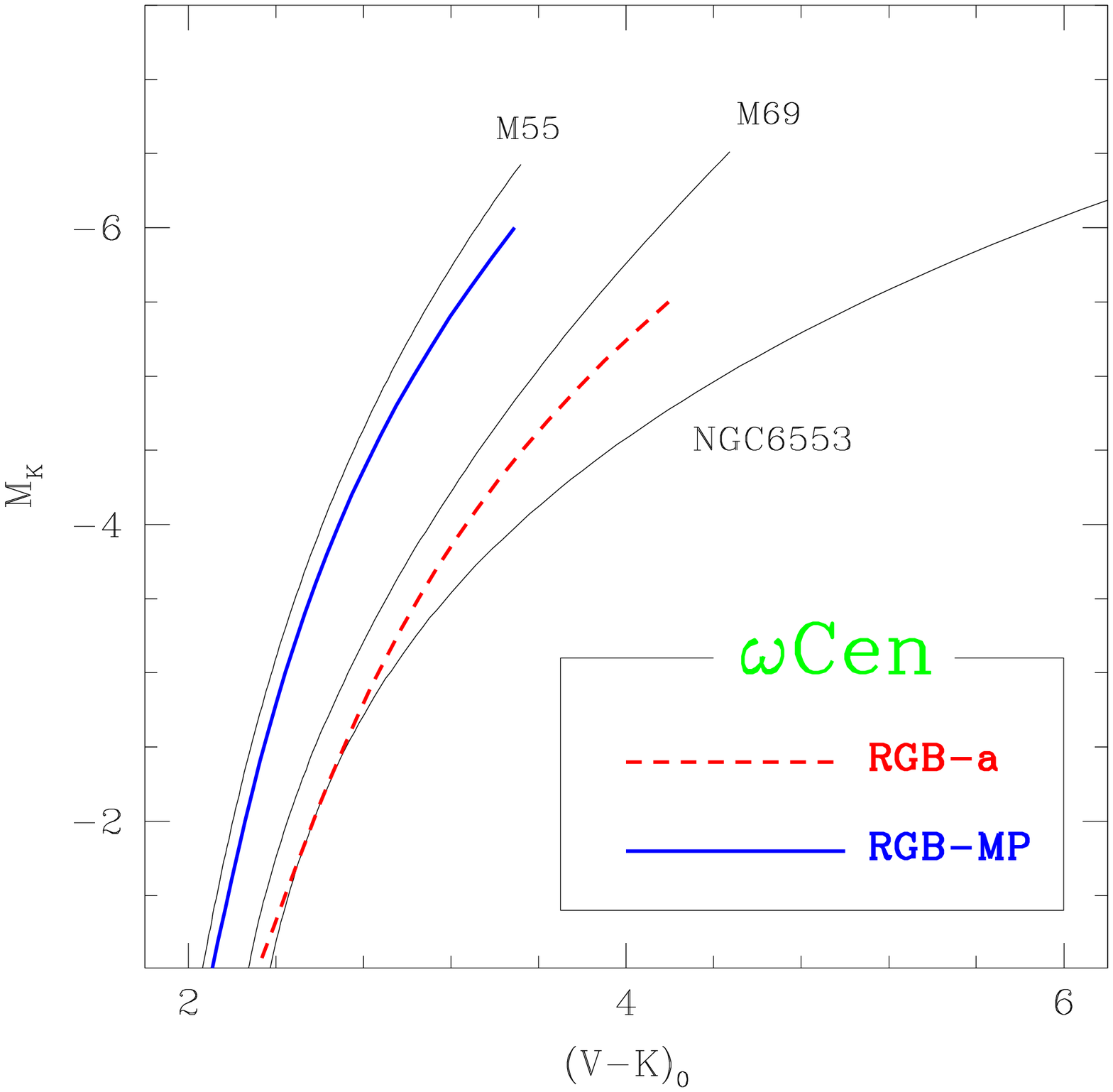}
\caption{{\it Left panel:} $(K,B-K)$-CMD for more than 50,000 stars
 in the central region of $\omega$ Cen. The RGB mean ridge line of the metal poor
  and the anomalous population are overplotted.
  {\it Right panel:} The mean ridge lines of the metal poor (solid line)
  and the anomalous population (dashed line)
  are compared with the RGB ridge lines of M55, M69
  and NGC6553 from F00.}
\end{figure}

\section{Mid-IR Observations of GGCs: probing the mass loss process
along the  Red Giant Branch}

 A complete, quantitative understanding of the physics of mass loss processes 
and the precise knowledge of the gas and dust content 
in GGCs is crucial in the study of Population 
II stellar systems and their impact on the Galaxy evolution.
Despite its importance, mass loss is  still a poorly
understood process.

In order to shed some light on  mass loss processes along the RGB 
  we performed (Origlia et al 2002) a deep   Mid-IR survey with ISOCAM
   of the very central regions 
of six, massive clusters: 47~Tuc, NGC~362,
 $\omega$~Cen, NGC~6388, M15 and  M54. Mid-IR observations
 are the ideal tool to study mass loss, since  they  could detect
  an outflowing gas fairly far away from the star
(typically, tens/few hundreds stellar radii). 
 
Two different
filters ([12], [9.6]) in the 10 $\mu $m spectral region have been used. 
The mid-IR colors  have been then combined with near-IR colors 
in order to obtain 
photometric indices  ($K-[12]$ or $K-[9.6]$), 
which are sensible  tracers of  circumstellar dust excess.
Figure~3 shows the $M_{\rm bol},~(J-K)_0$ and $M_{\rm bol}
,~(K-[12])_0$ CMDs.
Stars with $(K-[12])_0\ge 0.65$ are classified as 
sources with significant 
dust excess and are marked with filled symbols in the Figure.

There are  a series of interesting  results suggested by this Figure:
{\it (i)} all the stars showing evidence of mid-IR circumstellar
dust excess are in the upper 1.5 bolometric
magnitudes of the RGB, suggesting that
{\it significant mass loss occurs only at the very end of the RGB 
evolutionary stage};
{\it (ii)} only
20 out of the 52 ($\sim 40\%$) ISOCAM 
sources detected in the upper 1.5 bolometric
magnitudes of the RGB ($M_{\rm bol}\le-2.5$) show   evidence of   circumstellar
dust excess; by correcting
for stars not detected because of the low spatial resolution of
ISOCAM, dusty envelopes are inferred around about 15\% of the brightest 
giants, this suggests that {\it the mass loss
process is episodic}; {\it (iii)}~{\it there is no evidence of
any dependence of 
mass loss occurrence  on the cluster metallicity}.

\begin{figure}
\vspace{-0.in}
\plotfiddle{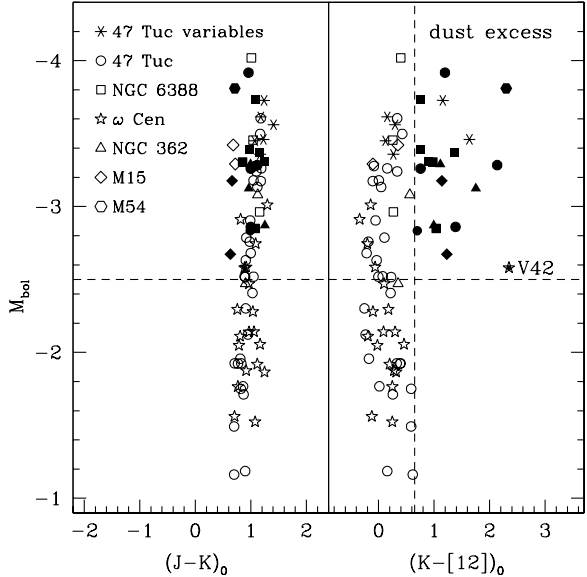}{7cm}{0}{40}{40}{-140}{-60} 
\caption{ $M_{\rm bol}$,$~(J-K)_0$ (left panel) and 
$M_{\rm bol}$,$~(K-[12])_0$ 
(right panel) de-reddened  CMD   of the ISOCAM point sources 
detected in  six globular clusters (from Origlia et al 2002).
Sources with $(K-[12])_0\ge 0.65$ (dashed vertical line)
are classified as sources with significant 
dust excess and are marked with filled symbols.}
\end{figure}

\section{UV observations: probing the hot stellar population in GGCs}

 Although the  CMD of an old stellar population (as a GGC) is dominated,
 in the {\it classical} $(V,B-V)$-plane,  
 by the cool
 stellar component, relatively populous hot stellar components do
  exist in  GGCs and are  strong emitters in the UV
 (hot post-Asymptotic Giant Branch stars, 
 blue HB, BSS, various by-products of binary 
 system evolution, and so on). 

The advent
of the Hubble Space Telescope (HST),  whith its unprecedented
 spatial resolution and 
imaging/spectroscopic capabilities in the UV, has
 given a new impulse to the study of hot stars in GGCs.
 We are involved in a 
 long-term observational programme which uses
 HST to perform UV observations
 in  a selected sample of GGCs.
 In this section I summarize   the most recent results obtained
 for BSS (a few additional results on the search of peculiar
 objects can be found in
 the poster contribution by Sabbi et al. and Ferraro et al.
 in this book).  

\begin{figure}
\vspace{-0.in}
\plotone{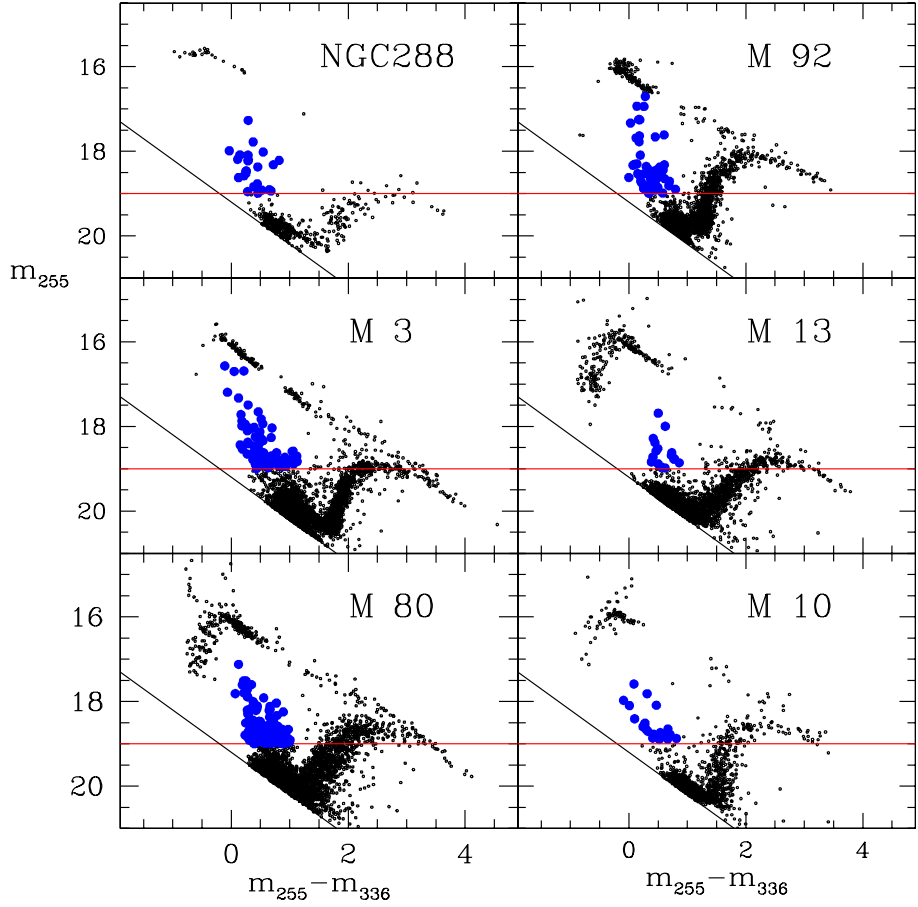} 
\caption{ 
($m_{255}, m_{255}-m_{336}$) CMDs for the 6  clusters observed with HST. 
Horizontal and vertical shifts have been applied to all CMDs in order  
to match the main sequences of M3.  
The horizontal solid line corresponds to $m_{255}=19$. 
The bright BSS candidates are marked as large filled circles
(from Ferraro et al 2002).
}
\end{figure} 
  
\subsection{Blue Straggler Stars in the UV} 

Blue Straggler stars (BSS), first discovered by Sandage (1953) in M3,
  are commonly defined as   stars  
  brighter and bluer (hotter) than the main sequence (MS) turnoff
(TO), lying along an apparent extension of the MS, and thus
mimicking a rejuvenated stellar population. The existence of such a
population has been a puzzle for many years, and even now its
formation mechanism is not completely understood, yet. At present,
the leading explanations involve mass transfer between binary
companions or the merger of a binary star system or the collision of
stars (whether or not in a binary system). Direct measurements
(Shara et al. 1997; Gilliland et al. 1998) 
and indirect evidence have in fact shown that BSS are
more massive than the normal MS stars, pointing again toward a
collision or merger of stars. Thus, the BSS represent the link
between classical stellar evolution and dynamical processes
(see Bailyn 1995). The
realization that BSS are the ideal diagnostic tool for a
quantitative evaluation of the   dynamical interaction effects
inside star clusters has led to a remarkable burst of searches and
systematic studies, using UV and optical broad-band photometry.

\begin{figure}
\vspace{-0.in}
\plotfiddle{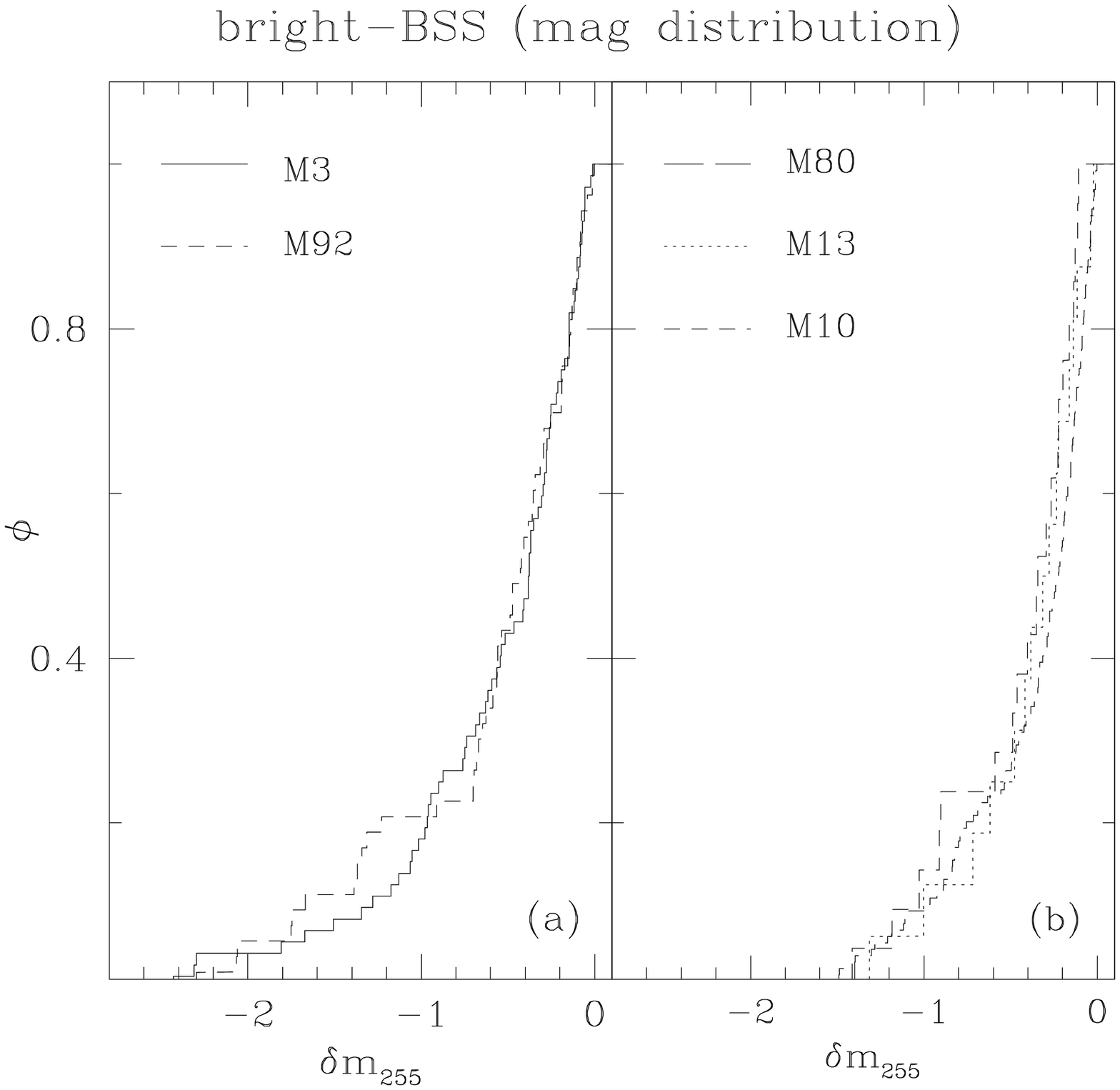}{6.5cm}{0}{38}{40}{-140}{-70}  
\caption{ 
 Cumulative  magnitude distributions for the {\it bright} 
 BSS   for each of the  
 six clusters. The $\delta m_{255}$ parameter is the b-BSS
   magnitude with respect to the  threshold 
  ($m_{255}=19$). 
 In {\it Panel (a)} the BSS distributions 
 for  M3 and M92 (the two clusters for which the  
 BSS distribution extends up to more than  
 two magnitudes brighter than the threshold) are compared. 
 In  {\it Panel (b)} the BSS magnitude distributions for the  other 3 
  clusters are plotted.}
\end{figure}

Our group has actively participated to this extensive surveys and
has published some of the first and most complete catalogs of BSS
in GCs (Fusi Pecci et al 1992; Ferraro,Bellazzini \& Fusi Pecci
1995; Ferraro et al 2002). 
These works have significantly contributed to form the nowadays
commonly accepted idea that BSS are indeed a normal component of
stellar populations in clusters, since they are present in all of
the properly observed GGCs. However, according to Fusi Pecci et al. 
(1992) BSS in different environments could have different origin.
In particular, BSS in loose GGCs might be produced from coalescence
of primordial binaries, while in high density GGCs (depending on
survival-destruction rates for primordial binaries) BSS might arise
mostly from stellar interactions, particularly those which involve
binaries. Thus, while the suggested mechanisms for BSS formation
could be at work in clusters with different environments (Ferraro,
Bellazzini, \& Fusi Pecci, 1995; Ferraro et al. 1999) 
there is evidence that they could
also act simultaneously within the same cluster (as in the case of M3,
see Ferraro et al.
1993; Ferraro et al. 1997). 
Moreover, as shown by Ferraro et al. (2002), both
the BSS formation channels (primordial binary coalescence and
stellar interactions) seem to be equally efficient in producing BSS
in different environments, since the two clusters that show the largest
known BSS specific frequency, i.e. NGC~288 (Bellazzini et al. 2002)
and M~80 (Ferraro et al. 1999), represent two exteme cases of
central density concentration among the GGCs ($Log \rho_0=2.1$ and
$5.8$).
Particularly interesting is the case of M80
which shows an exceptionally high BSS content: more 
than 300 BSS have been discovered in M80 (Ferraro et al 1999). 
This is {\it the largest and most 
   concentrated BSS population ever found in a GGC}. 
   Since M80 is the  GGC which has the largest central density among those 
   not yet core-collapsed, this discovery could be the first direct evidence 
   that stellar collisions could indeed be effective in delaying the core 
   collapse.

Figure 4 shows the ($m_{255},~m_{255}-m_{336}$) CMDs for   six
clusters observed in the UV with HST (Ferraro et al 2002).
 More than 50,000 stars are plotted in the six panels of
Figure 4. The CMD of each cluster has been shifted to 
match that of M3  using
the brightest portion of the HB as the normalization
region. The solid
horizontal line (at $m_{255}=19$) in the figure shows the 
 threshold
magnitude for the selection of bright (hereafter bBSS) sample.
Such a dataset allows a direct comparison of the  
photometric properties of bBSS in different clusters. In particular,
we have found   evidence (Ferraro et al. 2002) 
for a possible connection between
the presence of a blue tail in the HB and the BSS UV-magnitude
distribution: GGCs without HB blue tails have BSS-Luminosity
Function (LF) extending to
brighter UV magnitudes with respect to GGCs with blue tails.

In Figure 5 the magnitude distributions (equivalent to a LF) 
of bBSS for the six clusters are compared.  In doing
this we use the parameter $\delta m_{255}$ defined as the magnitude of
each bBSS (after the alignment showed in Figure 4) with respect to the
magnitude threshold (assumed at $m_{255}=19$ - see Figure 4). Then
$\delta m_{255}= m^{bBSS}_{255}-19.0$. From the comparison shown in
Figure 5 ({\it panel(a)}) the bBSS magnitude distributions for M3 and
M92 appear to be quite similar and both are significantly different from
those obtained in the other clusters. This is essentially because in
both clusters the bBSS magnitude distribution seems to have
a tail extending to brighter magnitudes (the bBSS magnitude tip
reaches $\delta m_{255}\sim -2.5$). A KS test applied to these two
distributions yields a probability of $93\%$ that they are extracted
from the same distribution. In {\it panel(b)} we see that the bBSS
magnitude distribution of M13, M10 and M80 are essentially
indistinguishable from each other and significantly different from M3
and M92. A KS test applied to the three LFs confirms that they
are extracted from the same parent distribution. Moreover, a KS test
applied to the total LFs obtained by combining the data for the two
groups: M3 and M92 ({\it group(a)}), and M13, M80 and M10 ({\it
group(b)}) shows that the the bBSS-LFs of {\it group(a)} and 
{\it group(b)} are
not compatible (at $3\sigma$ level).

It is interesting to note that the clusters grouped on the basis of
bBSS-LFs have some similarities in their HB morphology. The three
clusters of {\it group(b)} have an extended HB blue tail; the two clusters
of {\it group(a)} have no HB extention. Could there be a connection between
the bBSS photometric properties and the HB morphology? This
possibility needs to be further investigated.

\acknowledgments

It is a pleasure to thank  all the collaborators involved in this 
vast project. In particular, I want to thank Elena Pancino
for a critical reading of this manuscript and
Livia Origlia for her continuos support.
The financial support of 
the {\it Agenzia Spaziale Italiana} (ASI) and  of the {\it Ministero della
Istruzione dell'Universit\`a e della Ricerca} (MIUR) is kindly acknowledged.

\end{document}